# Incorporating action and reaction into a particle interpretation for quantum mechanics – Dirac case

Roderick I. Sutherland

Centre for Time, University of Sydney, NSW 2006 Australia

rod.sutherland@sydney.edu.au

**Abstract**

A weakness which has previously seemed unavoidable in particle interpretations of quantum mechanics (such as in the de Broglie-Bohm model) is addressed here and a resolution proposed. The weakness in question is the lack of action and reaction occurring between the model's field (or "pilot wave") and the particle. Although the field acts on the particle, the particle does not act back on the field. It is shown here that this rather artificial feature is, in fact, not necessary and can be fully eliminated while remaining consistent with the usual quantum predictions. Mathematically this amounts to demonstrating that there exists a suitable Lagrangian density function which generates equations coinciding with quantum mechanics yet incorporates the desired action and reaction. As a by-product, an appealing explanation emerges to another long-standing question, namely why the mathematical formalism of quantum mechanics seems only to be describing fields when measurements generally detect localised particles. A further bonus is that the hitherto unrelated concept of a gauge transformation is found to arise naturally as an essential part of the formalism. In particular, the phase S of the gauge transformation is seen to be the action function describing the hidden motion of the particle.

## 1. Introduction

This paper relates to interpretations of quantum mechanics which take the underlying reality to consist of particles having definite trajectories. The de Broglie-Bohm model [1,2] is the basic and best-known example of this type, but extended versions with arguably more appealing characteristics can and have been constructed by pursuing a Lagrangian approach [3,4,5]. Such later versions offer, for example, the possibility of restoring conservation of energy and momentum and the possibility of a source for the (otherwise mysterious) field guiding the particle. Nevertheless, in order to ensure that such models remain consistent with quantum mechanics, it has so far seemed necessary to retain the rather artificial feature that the particle does not act back on the field. It will now be shown here, however, that this lack of two-way interaction is, in fact, not necessary and can be rectified without contradicting the usual quantum predictions. This will be demonstrated by considering the Dirac equation as an example and employing a suitable Lagrangian formulation.

As a result of introducing this refinement to the usual de Broglie-Bohm picture, three related advantages emerge. First, a possible answer arises to the long-standing question as to why the mathematical formalism seems to be concerned only with fields when it is particles which are generally observed in experiments. It is found that there are actually two field equations which can be obtained from the Lagrangian formalism presented here. One of these equations contains both particle and field terms and gives a full description of individual events. It is, however, non-linear and would be difficult to solve. The other equation, a statistical version, is obtained by acknowledging that the particle's position is not precisely known and is instead distributed via the relativistic form of the usual Born rule. This yields an equation which is then recognised as the usual quantum wave equation (in the present case, the Dirac equation), thereby recovering all the established predictions of quantum mechanics. In taking this statistical step, however, the particle terms are seen to be lost in the process, making it appear as if fields are the only physical reality involved.

The second welcome feature to emerge concerns the historical fact that the de Broglie-Bohm approach can be viewed as being closely related to the Hamilton-Jacobi formulation of classical mechanics. In particular, the momentum **p** of the particle in the original model is related to the phase S of the wavefunction via the simple and elegant relationship:

$$\mathbf{p} = \frac{\nabla S}{\hbar} \tag{1}$$

This equation was introduced in the Schrodinger case but a subsequent, prevalent view has been that it cannot be employed for particles with spins other than zero. In the approach presented here, however, a Hamilton-Jacobi formulation incorporating the relativistic version of Eq. (1) arises naturally, even though it is now spin-half being considered and S is no longer simply the phase of the wavefunction.

The third point is that the model provides an unexpected physical link to the concept of a gauge transformation, which normally stands quite separately. On this point, the gradient of the phase associated with such a transformation is necessarily identified here as being the generalised momentum of the underlying particle.

The structure of the paper is as follows. Sections 2 to 4 summarise the basic Lagrangian formalism needed to accommodate action and reaction within quantum mechanics. A general statistical framework is then introduced in Sec. 5 and a particular probability density is postulated in Sec. 6. Agreement with the standard wave equation is then demonstrated in Sec. 7. The conservation laws corresponding to this action/reaction picture are formulated and shown to hold in Sec. 8 by introducing the energy-momentum tensor associated with the particle/field system and confirming that it has zero divergence. Finally, the extension to the many-particle case is discussed briefly in Sec. 9, where two possible ways of formulating such a generalisation are described and compared.

**2. Appropriate Lagrangian formalism**

It is sufficient to consider the single-particle case in order to illustrate the essential idea. The treatment here will be relativistic with the units chosen such that $\hbar = c = 1$ for simplicity. Pursuing a similar strategy to the author's previous papers [3,4], the discussion will proceed by analogy with the well-known electromagnetic formalism. As outlined in various textbooks (e.g., [6,7]), the classical description of a charged particle interacting with an electromagnetic field can be summarised by an overall Lagrangian density $\mathcal{L}$ for the field and particle combined. This expression has the general form:

$$\mathcal{L} = \mathcal{L}_{\text{field}} + \mathcal{L}_{\text{particle}} + \mathcal{L}_{\text{interaction}} \tag{2}$$

In terms of the particle's 4-velocity $u^\alpha$ and the electromagnetic 4-potential $A^\alpha$, the above equation can be written more explicitly in the form[1]:

$$\mathcal{L} = \mathcal{L}_{\text{field}} - \sigma_0 m (u_\alpha u^\alpha)^{\frac{1}{2}} - \sigma_0 q\, u_\alpha A^\alpha \qquad (\alpha = 0,1,2,3) \qquad (3)$$

The quantity $\sigma_0$ here is the rest density distribution of the particle through space, which involves a delta function because the particle's "matter density" is concentrated at one point. As demonstrated in the previous work, a Lagrangian density similar to this one but with $A^\alpha$ replaced by a different 4-vector (one related to the wavefunction) is capable of reproducing the standard quantum mechanical results. In the case to be considered here, this new 4-vector takes the form of the 4-current density $j^\alpha$ associated with the Dirac equation. By analogy with the electromagnetic case in Eq. (3), the postulated Lagrangian density is then chosen to be:

$$\mathcal{L} = \mathcal{L}_{\text{field}} - \sigma_0 k \rho_0 (u_\alpha u^\alpha)^{\frac{1}{2}} - \sigma_0 k u_\alpha j^\alpha \qquad (4)$$

Here the term $\mathcal{L}_{\text{field}}$ is now taken to be the usual textbook form of the Lagrangian density describing the wavefunction alone, k is an arbitrary constant [2] and $\rho_0$ is the magnitude of the 4-current density:

$$\rho_0 = (j_\alpha j^\alpha)^{\frac{1}{2}} \qquad (5)$$

An obvious novel feature here is that the rest mass m in Eq. (3) has been replaced by the quantity $k\rho_0$ in Eq. (4). This is the key point which allows a model that agrees with quantum mechanics to be constructed. An additional point to note is that, in contrast to the author's previous work, the sign of the last term in Eq. (4) has been changed to negative. It is this step which now allows a model incorporating action and reaction to be constructed for the Dirac case.

It will be shown over the next few sections how the proposed Lagrangian density in Eq. (4) leads to the usual wave equation.

---

[1] Here the term $\mathcal{L}_{\text{field}}$ can be expressed in further detail as a function of the derivatives of the 4-vector $A^\alpha$. The quantities m and q are the particle's rest mass and charge, respectively. Eq. (3) has been written in manifestly Lorentz covariant form and it is assumed that there is a summation over any repeated index and that the metric tensor has signature $(+---)$.

[2] This constant must have dimensions $ML^3$ in order to balance the units. In previous work, its value was set equal to one for simplicity. If, instead, one were to consider building it out of the known physical constants involved in quantum mechanics, dimensional analysis would give the possible choice $\dfrac{\hbar^3}{m^2 c^3}$, which for, e.g., an electron has the value $1.3 \times 10^{-65} \text{kg m}^3$.

## 3. Field equation

For the Dirac case[3], the field term to be inserted in Eq. (4) is the following standard expression from text books[4]:

$$\mathscr{L}_{\text{field}} = \tfrac{1}{2}\left[ i\,\overline{\phi}\gamma^\beta(\partial_\beta \phi) - i(\partial_\beta \overline{\phi})\gamma^\beta \phi \right] - m\,\overline{\phi}\phi \tag{6}$$

and the 4-current density $j^\alpha$ has the form:

$$j^\alpha = \overline{\phi}\gamma^\alpha \phi \tag{7}$$

The letter $\phi$ has been used here rather than the usual $\psi$ because the field $\phi$ acting on the particle in this model is not directly equivalent to the wavefunction at this stage. The connection between these two functions will emerge shortly.

A field equation for $\phi$ can be obtained from the Lagrangian density (4) by performing a variation of $\overline{\phi}$. This standard procedure is carried out in Appendix 1 and yields the following result[5]:

$$i\gamma^\alpha \partial_\alpha \phi - m\phi = \sigma_0 k \left( u_\alpha + \frac{j_\alpha}{\rho_0} \right)\gamma^\alpha \phi \tag{8}$$

The left hand side of this equation is seen to consist of the usual Dirac terms, whereas the right hand side can be viewed as a source term arising from the (hidden, but continuously existing) particle. In this regard it should be kept in mind that the quantity $\sigma_0$ here contains a delta function representing the position of the particle. Also, to avoid confusion it needs to be mentioned that the sign in the bracket is intentionally different compared with the author's previous model. It is, of course, necessary to explain why the right hand term in this equation is not ruled out immediately by the existing experimental evidence.

In previous work, the field equation (8) was simply reduced to the standard Dirac equation by making the extra assumption that the bracket on the right hand side was zero. Since the sign in the bracket was negative in that work, this simply meant restricting the particle's 4-velocity to:

$$u_\alpha = \frac{j_\alpha}{\rho_0} \tag{9}$$

---

[3] In the following equations, $\phi(x)$ is a spinor field and $\overline{\phi}(x)$ is its adjoint, both being functions of position and time: $x \equiv (x^0, x^1, x^2, x^3)$. The symbol $\partial_\alpha$ is an abbreviation for the partial derivative $\dfrac{\partial}{\partial x^\alpha}$, whilst $\gamma^\alpha$ represents the Dirac matrices and m is the rest mass of the particle in question.

[4] The sign of $\mathscr{L}_{\text{field}}$ has been changed here compared with the author's previous work for consistency with most texts (e.g., [8]).

[5] For simplicity, Eqs. (6) and (8) have been limited to the free-space case and do not contain any external potentials. A term representing an external 4-vector potential $A^\alpha$ can, however, easily be added to each. For example, in Eq. (6) it would be of the form $A_\alpha j^\alpha$, where $j^\alpha$ is given by Eq. (7).

which is the same as the guidance equation of the de Broglie-Bohm model once $\phi$ is assumed to be the usual wavefunction. Here the same result could, of course, be achieved by choosing:

$$u_\alpha = -\frac{j_\alpha}{\rho_0} \quad (10)$$

Either way, however, eliminating the source term from the field equation in this fashion also has the effect of removing any influence of the particle on the field, thereby ruling out any possibility of two-way interaction. Fortunately there is an alternative way of recovering the standard Dirac equation from the field equation (8) without this negative consequence. This other way of proceeding becomes available in the statistical case once it is acknowledged that the particle's position is not known precisely and needs to be described by a probability distribution. This approach will be developed from Sec. 5 onwards. As a preliminary step, however, the generalised momentum of the particle will now be derived so as to write Eq. (8) in a more convenient form.

## 4. Generalised momentum

The overall Lagrangian density in Eq. (4) can be expressed equivalently as the following sum:

$$\mathcal{L} = \mathcal{L}_{field} + \sigma_0 L \quad (11)$$

where L is the Lagrangian (as opposed to a Lagrangian density) governing the motion of the particle. Comparing Eqs. (4) and (11), the expression for L is:

$$L = -k\left[\rho_0(u_\alpha u^\alpha)^{1/2} + u_\alpha j^\alpha\right] \quad (12)$$

From this Lagrangian the particle's generalised momentum $p^\alpha$ can be derived via the following standard formula from classical mechanics[6]:

$$p^\alpha = -\frac{\partial L}{\partial u_\alpha} \quad (13)$$

which, as shown in Appendix 2, leads to the following result:

$$p^\alpha = k(\rho_0 u^\alpha + j^\alpha) \quad (14)$$

This generalised momentum is also expressible via the following familiar Hamilton-Jacobi expression:

$$p^\alpha = -\partial^\alpha S \quad (15)$$

where S is the action defined by:

$$S(x) = \int_{t_0}^{t} L \, d\tau \quad (16)$$

---

[6] Here and in Eq. (15) the signs have been chosen to be consistent with the usual textbook definitions $p^i = \partial L / \partial u^i$ (i = 1, 2, 3) and $\mathbf{p} = \nabla S$, respectively, for the spatial components of this momentum.

This last equation can be viewed as expressing the action integral for the particle's actual path as a function of the coordinates $x = (t, \mathbf{x})$ at the upper limit of integration. Here L is again the Lagrangian given in Eq. (12), $\tau$ is the proper time along the particle's world line and $t_0$ is an arbitrary time in the past at which the particle's position $\mathbf{x}_0$ is taken to be fixed. Eqs. (14) and (15) can now be combined to give:

$$\partial^\alpha S = -k(\rho_0 u^\alpha + j^\alpha) \qquad (17)$$

and inserting this result back into the field equation (8) yields:

$$i\gamma^\alpha \partial_\alpha \phi - m\phi = -\frac{\sigma_0}{\rho_0}(\partial_\alpha S)\gamma^\alpha \phi \qquad (18)$$

This form of the field equation will be more suitable for present requirements.

## 5. Statistical framework

It will now be pointed out with the help of a well-known thought experiment that the usual wavefunction of quantum mechanics must be connected in only a statistical way to the field required in this model. Consider a point source emitting particles isotropically, so that each particle's wavefunction will evolve away from the source in a spherically symmetric fashion. If it is assumed that there is also a shell-like detector surrounding the point source at a certain radial distance, each particle will eventually be detected and will be seen to have travelled in its own particular direction, so that each particle breaks the spherical symmetry. Now the model proposed here entails that the particle is acting as a source of the field. It is therefore to be expected that the field, unlike the wavefunction, will be greater in the vicinity and direction of the path which the particle actually takes. This fact that the wavefunction will expand symmetrically but the field will not indicates that the two quantities can only be related statistically.

The statistical element which this argument requires will now be introduced by returning to the overall Lagrangian density (4) and taking a weighted average over the possible positions of the particle. The specific form of the position probability distribution is not immediately important and will be postponed until Sec. 6. In carrying out this averaging process, the explicit expression for the rest density distribution $\sigma_0$ in Eq. (4) will now be needed. This quantity is known to have the following Lorentz covariant form [3]:

$$\sigma_0 = \frac{1}{u^0}\delta^3[\mathbf{x} - \mathbf{x}_p(\tau)] \qquad (19)$$

where $\mathbf{x}_p$ is the particle's spatial position as a function of proper time $\tau$ and $\mathbf{x}$ is an arbitrary point in space. The desired weighted average can be obtained by multiplying the Lagrangian density (4) by the (as yet unknown) probability distribution $P(\mathbf{x}_p)$ and then integrating over $\mathbf{x}_p$. It will be assumed that the integral of $P(\mathbf{x}_p)$ over all space is equal to one. With the $\mathbf{x}$'s and $\mathbf{x}_p$'s displayed explicitly, the following statistical version of the Lagrangian density is obtained[7]:

$$\bar{\mathcal{L}}(\mathbf{x}) = \iiint_{-\infty}^{+\infty} \left\{ \mathcal{L}_{field}(\mathbf{x}) + \frac{1}{u^0} \delta^3(\mathbf{x} - \mathbf{x}_p) \left[ -k\rho_0(\mathbf{x})(u_\alpha u^\alpha)^{\frac{1}{2}} - k u_\alpha j^\alpha(\mathbf{x}) \right] \right\} P(\mathbf{x}_p) \, d^3 x_p \tag{20}$$

Performing the integrals then yields:

$$\bar{\mathcal{L}}(\mathbf{x}) = \mathcal{L}_{field}(\mathbf{x}) + \frac{P(\mathbf{x})}{u^0} \left[ -k\rho_0(\mathbf{x})(u_\alpha u^\alpha)^{\frac{1}{2}} - k u_\alpha j^\alpha(\mathbf{x}) \right] \tag{21}$$

Having obtained this more general expression, the next step is to find the field equation which it implies. Since the quantities $P(\mathbf{x})$ and $u^0$ are independent of the field for the purposes of this derivation, the steps involved are essentially the same as those already carried out in Appendix 1, the only change being the replacement of $\sigma_0$ by $\frac{P(\mathbf{x})}{u^0}$. Therefore, by analogy with Eq. (18) earlier, the following result can be stated for the statistical version of the field equation:

$$i\gamma^\alpha \partial_\alpha \psi - m\psi = -\frac{P(\mathbf{x})}{\rho_0 u^0} (\partial_\alpha S) \gamma^\alpha \psi \tag{22}$$

The different letter $\psi$ has been used here rather than $\phi$ to highlight the fact that the field solution will now differ from that in previous sections because it is a solution of Eq. (22) instead of Eq. (18). The earlier solution is the actual field interacting with the particle whereas the new solution is the result obtained when only a probability distribution is inserted for the source particle's position, rather than a definite value[8].

## 6. Postulated probability distribution

In order to demonstrate consistency with standard quantum mechanics, a specific expression is now needed for the probability distribution $P(\mathbf{x})$. Such an expression cannot be derived and must be postulated separately, being more akin to a boundary condition than a basic part of the mathematical structure. It should, however, satisfy certain desirable conditions such as being conserved and being positive definite. An argument can be advanced in favour of a particular choice and, not surprisingly, this choice turns out to be the relativistic version of the Born rule.

---

[7] Most terms in this equation are functions of time as well but, in the interests of notational simplicity, this detail has not been shown.
[8] Note that the quantities $\rho_0$ and $\partial_\alpha S$ in Eq. (22) remain defined by the expressions in Eqs. (5) and (17) but with these expressions now written in terms of the latter solution $\psi$.

Conservation of probability at each point in space requires the distribution to satisfy a continuity equation and such a relationship can easily be derived from the field equation (22). Specifically, using the adjoint equation[9] to (22):

$$-i\partial_\alpha \bar{\psi}\gamma^\alpha - m\bar{\psi} = -\frac{P(\mathbf{x})}{\rho_0 u^0}(\partial_\alpha S)\bar{\psi}\gamma^\alpha \qquad (23)$$

the desired continuity equation is obtained by the familiar method of multiplying (22) on the left by $\bar{\psi}$ and (23) on the right by $\psi$ then subtracting the resulting two equations to yield the following standard Dirac result:

$$\partial_\alpha(\bar{\psi}\gamma^\alpha\psi) = 0 \qquad (24)$$

The point here is that the source terms in Eqs. (22) and (23) have cancelled out and so this result is obtained regardless of the choice of probability distribution. The expression in the bracket of Eq. (24) has the form of the usual Dirac 4-current density (now expressed in terms of the $\psi$ solution rather than $\phi$) and Eq. (24) ensures that it remains conserved. In addition, the zeroth component of this expression is positive definite and is therefore a candidate for $P(\mathbf{x})$.

It will therefore be postulated here that the particle's position should be described statistically by the current density 4-vector already employed earlier for other purposes, viz.:

$$j^\alpha = \bar{\psi}\gamma^\alpha\psi \qquad (25)$$

and hence that the appropriate expression for $P(\mathbf{x})$ is given by the zeroth component of $j^\alpha$:

$$P(\mathbf{x}) = \bar{\psi}\gamma^0\psi \qquad (26)$$

## 7. Reduction to the Dirac equation

The Hamilton-Jacobi formalism provides a specific solution for $u^\alpha$. The particular solution corresponding to the Lagrangian density (21) will now be introduced by returning to the generalised momentum relationship (17) and rearranging it into the form:

$$u^\alpha = -\frac{\partial^\alpha S(x) + kj^\alpha(x)}{k\rho_0(x)} \qquad (27)$$

with the understanding that the quantities on the right hand side are now functions of the new solution $\psi(x)$. From Eq. (27) it is clear that $u^\alpha$ is a function of x, as is usual in a Hamilton-Jacobi formulation. Therefore this equation implies there will be only a single value of $u^\alpha$ for each position[10], which then allows the 4-current density $j^\alpha$ to be written in the following product form[11]:

$$j^\alpha = \rho_0 u^\alpha \tag{28}$$

with $\rho_0$ given by Eq. (5) as usual. The corresponding probability density for the particle's position is:

$$P(\mathbf{x}) = \rho_0 u^0 \tag{29}$$

---

[9] In taking the adjoint, note that the quantities $\rho_0$ and $\partial^\alpha S$ are real. This follows because $j^\alpha$ as defined in Eq. (7) is real and $\rho_0$ and $\partial^\alpha S$ can then be expressed in terms of $j^\alpha$ via Eqs. (5) and (17), respectively.

Inserting this result into Eq. (22), the field equation then reduces to:

$$i\gamma^\alpha \partial_\alpha \psi - m\psi = -(\partial_\alpha S)\gamma^\alpha \psi \tag{30}$$

Although this simplified version of the field equation still seems different from the standard Dirac equation because of the right hand term, a final step will clarify the situation. This entails switching to a new field quantity $\Psi(x)$ via the following change of notation[12]:

$$\Psi(x) = \psi(x) e^{-iS(x)} \tag{31}$$

which is akin to performing a gauge transformation. The quantity $S(x)$ here is the same action as defined earlier and the function $\Psi(x)$ will shortly be identified with the Dirac wavefunction. Note that S is real but the spinors $\psi$ and $\Psi$ will generally be complex. Under this change of notation the field equation (30) becomes:

$$i\gamma^\alpha \partial_\alpha (\Psi e^{iS}) - m\Psi e^{iS} = -(\partial_\alpha S)\gamma^\alpha \Psi e^{iS} \tag{32}$$

Carrying out the derivative in the 1st term and then cancelling the factors $e^{iS}$, the result is seen to reduce to:

$$i\gamma^\alpha \partial_\alpha \Psi - m\Psi = 0 \tag{33}$$

which is just the standard Dirac equation. It has therefore been shown that the original Lagrangian density (4) leads, via a statistical treatment together with a change of notation, to the correct wave equation in the Dirac case[13].

It is interesting to compare the relative advantages of the different field equations that have been derived here. The "non-statistical" Eq. (8) clearly contains both a field quantity $\phi$ and particle quantities $\sigma_0$ and $u^\alpha$ while giving a precise description of the influence (via the source terms) of the particle on the field $\phi$. This equation is non-linear, however, and would be difficult to solve. It would also need to be solved simultaneously with the particle's equation of motion. Turning to the "statistical" case of Eq. (33), which is just the standard Dirac equation, suddenly the equation is linear and can be solved relatively easily for $\Psi$ to obtain the usual quantum predictions. On the other hand, the price paid is that all evidence of a localised particle has been washed out of the equation, apparently indicating that fields are the only things which exist.

---

[10] Note that this $u^\alpha(x)$ relates to the statistical case described by Eq. (21) and (22) and so it is not the particle's actual 4-velocity but merely the value calculated once there is only a statistical input. Nevertheless, this value is uniquely determined at each position.

[11] Note that this step would not be possible if there were a range of possible $u^\alpha$ values at each x, since then the spatial components of $j^\alpha(x)$ could at most be decomposed into: $j^i = \rho(x) <v^i>_x$ $(i = 1,2,3)$, where $<v^i>_x$ is the mean value of the 3-velocity at x.

[12] This equation would actually have the form $\Psi = \psi e^{-iS/\hbar}$ if units yielding $\hbar = 1$ had not been chosen. There is no restriction imposed on S being larger than $\hbar$ and therefore there is the possibility that multiple S's could correspond to the same $\Psi$, but this does not affect the conclusions drawn here.

[13] Analogous results can be derived for the Schrodinger and Klein-Gordon cases.

Up to this point the focus has been on the free-space case for simplicity. It is important, however, to highlight the further possibility which would arise if an external 4-potential $A^\alpha$ were included in the Dirac equation. In that more general case, the extra term in Eq. (30) could disposed of by the alternative procedure of absorbing it into the 4-potential via a gauge-like transformation, i.e., by a change of notation for the 4-potential while keeping the lower case $\psi$ as the wavefunction.

Note that under the change of notation $\Psi = \psi e^{-iS}$, Eq. (25) keeps the same form but with $j^\alpha$ now written in terms of the wavefunction $\Psi$ as:

$$j^\alpha = \bar{\Psi}\gamma^\alpha\Psi \tag{34}$$

in accordance with the usual Dirac formulation.

The above discussion has shown how the particle can be still be influencing the field despite the apparent absence of a source term in the standard wave equation. Furthermore, the continued presence of the effect in the other direction (i.e., field on particle) can be trivially confirmed by combining Eqs. (28) and (34) to obtain[14]:

$$u^\alpha = \frac{\bar{\Psi}\gamma^\alpha\Psi}{\rho_0} \tag{35}$$

from which it is clear that the field is influencing the particle's 4-velocity. Hence the influence is seen to be two-way.

## 8. Energy and momentum conservation

As the particle and its associated field mutually interact they will continually exchange energy and momentum. Also, since the Lagrangian density in Eq. (4) is not an explicit function of the coordinates $x^\alpha$ (i.e., it is symmetric under space and time displacements), Noether's theorem implies the existence of an energy-momentum tensor $T^{\alpha\beta}$ having zero 4-divergence for the particle/field system:

$$\partial_\beta T^{\alpha\beta} = 0 \tag{36}$$

---

[14] This is the same guidance equation as for the de Broglie-Bohm model. It is also possible to combine Eqs. (15), (27) and (28) and obtain the further relationship $p^\alpha = 2kj^\alpha$.

This condition ensures overall conservation of energy and momentum during interaction. For a Lagrangian density of the present type, namely one which can be expressed in the form of Eq. (11), a general expression for $T^{\alpha\beta}$ is available [9]. As detailed in Appendix 3, the overall energy-momentum tensor for this case can be written as:

$$T^{\alpha\beta} = T^{\alpha\beta}_{\text{field}} + T^{\alpha\beta}_{\text{particle}} + T^{\alpha\beta}_{\text{interaction}} \tag{37}$$

where, for the present Dirac particle/field system, the individual terms are given by[15]:

$$T^{\alpha\beta}_{\text{field}} = \tfrac{1}{2}i\left[\bar{\phi}\gamma^{\beta}(\partial^{\alpha}\phi) - (\partial^{\alpha}\bar{\phi})\gamma^{\beta}\phi\right] + g^{\alpha\beta}\frac{\sigma_0}{\rho_0}(\partial_\lambda S)\bar{\phi}\gamma^{\lambda}\phi \tag{38}$$

$$T^{\alpha\beta}_{\text{particle}} = \sigma_0 p^{\alpha} u^{\beta} \tag{39}$$

$$T^{\alpha\beta}_{\text{interaction}} = 0 \tag{40}$$

In Appendix 4 the overall 4-divergence of this energy-momentum tensor is shown to be zero, thereby confirming conservation. It is also shown that the separate 4-divergences $\partial_\beta T^{\alpha\beta}_{\text{field}}$ and $\partial_\beta T^{\alpha\beta}_{\text{particle}}$ are not zero, indicating that energy and momentum exchanges are occurring between the field and the particle.

## 9. Many-particle case

Two possible ways of generalising the above formalism to the many-particle case will now be outlined, with the full details available elsewhere.

The first and more well-known approach (e.g., [11]) simply involves describing all the particles by a single, overall wavefunction. In this case, a corresponding Lagrangian density and action would then need to be defined in 3n dimensional configuration space, as is normal in a Hamilton-Jacobi treatment of n mutually interacting particles. Unlike the classical case, however, the field (and therefore physical reality) would be relegated to configuration space as well, which is less satisfactory. Also, this approach requires a preferred frame of reference in order to be consistent with the nonlocality implied by Bell's theorem [12], thereby clashing with the spirit of (experimentally well-confirmed) special relativity.

An alternative approach to the many-particle case is to maintain special relativity without any preferred frame and to conclude from Bell's theorem that retrocausality (i.e., backwards-in-time effects) must therefore be involved [13-21]. As is shown elsewhere [4], invoking retrocausality has the advantage of allowing a separate wavefunction to be defined for each particle once they are no longer interacting with each other. This then allows the fields and therefore the physical description to be returned to spacetime rather than residing in a 3n dimensional space. The configuration space formalism remains mathematically useful in this approach, just as it does in classical mechanics, but without corresponding literally to physical reality. This second alternative is favoured by the present author.

---

[15] The tensor $T^{\mu\nu}$ defined here is actually the "canonical" energy-momentum tensor, which is not necessarily symmetric and hence does not necessarily conserve angular momentum. Techniques exist to symmetrise this tensor [10].

As can be seen from the above considerations, the transition to the many-particle case is not straightforward regardless of one's preferred picture of underlying reality and it is a matter of taste whether one prefers to violate special relativity or to invoke retrocausality.

## 10. Discussion and Conclusions

Working within the context of a particle interpretation of quantum mechanics, a specific model has been constructed which incorporates action and reaction between the particle and the guiding field for the Dirac case. This model thereby demonstrates that two-way interaction can be achieved without contradicting the existing quantum predictions. The process is seen to involve exchanges of energy and momentum which conform to the usual conservation laws. Unlike in the author's preceding model [3,4], the mutual interaction does not reduce to zero in the special case of the quantum limit and continues unabated, although it becomes hidden from sight when the standard formalism of quantum mechanics is used.

The model has resulted in three side benefits, viz. (i) it provides a possible explanation for why we seem to be dealing purely with propagating fields in the standard theory even though experiments generally detect particles, (ii) it shows that a simple Hamilton-Jacobi formulation still remains possible once spinor wavefunctions are involved, and (iii) it brings to light a connection with the previously unrelated concept of a gauge transformation which emerges here in a natural way from the formalism.

Although the model gives rise to the standard Dirac equation and so to the usual predictions, there is some potential for it to make predictions which go beyond quantum mechanics. This is because the extra feature introduced here of the particle influencing the field provides further scope for testable consequences to be devised.

**Acknowledgement**

The author wishes to thank Ken Wharton for his numerous valuable suggestions on this work.

**Appendix 1**

The wave equation for the field $\phi$ can be found most simply by applying the usual Lagrange formula [8], which here takes the form:

$$\partial_\alpha \frac{\partial \mathcal{L}}{\partial(\partial_\alpha \bar{\phi})} - \frac{\partial \mathcal{L}}{\partial \bar{\phi}} = 0 \tag{41}$$

Using expression (11):

$$\mathcal{L} = \mathcal{L}_{\text{field}} + \sigma_0 L \tag{42}$$

and substituting it into the above equation then yields:

$$\left[\partial_\alpha \frac{\partial}{\partial(\partial_\alpha \bar{\phi})} - \frac{\partial}{\partial \bar{\phi}}\right] \mathcal{L}_{\text{field}} = -\left[\partial_\alpha \frac{\partial}{\partial(\partial_\alpha \bar{\phi})} - \frac{\partial}{\partial \bar{\phi}}\right] \sigma_0 L \tag{43}$$

With the aid of the field part of the Lagrangian density given in Eq. (6), the left hand side of this equation becomes:

$$\left[\partial_\alpha \frac{\partial}{\partial(\partial_\alpha \bar\phi)} - \frac{\partial}{\partial \bar\phi}\right] \mathcal{L}_{field} = \left[\partial_\alpha \frac{\partial}{\partial(\partial_\alpha \bar\phi)} - \frac{\partial}{\partial \bar\phi}\right]\left\{\tfrac{1}{2}\left[i\bar\phi\gamma^\beta(\partial_\beta\phi) - i(\partial_\beta\bar\phi)\gamma^\beta\phi\right] - m\bar\phi\phi\right\}$$

$$= \partial_\alpha(-\tfrac{1}{2}i\,\delta^\alpha_\beta\gamma^\beta\phi) - \tfrac{1}{2}i\,\gamma^\beta\partial_\beta\phi + m\phi$$

$$= -i\gamma^\alpha\partial_\alpha\phi + m\phi \qquad (44)$$

Turning to the term on the right hand side of Eq. (43) and noting that the Lagrangian L in Eq. (12) is not a function of $\partial_\alpha\bar\phi$, this right hand term reduces to:

$$\sigma_0 \frac{\partial L}{\partial \bar\phi} \qquad (45)$$

Now, with the aid of Eq. (5) together with the identity $(u_\alpha u^\alpha)^{1/2} = 1$, Eq. (12) can be written as:

$$L = -k\left[(j_\alpha j^\alpha)^{1/2} + u_\alpha j^\alpha\right] \qquad (46)$$

Since this expression depends on $\bar\phi$ only via $j^\alpha$, it is more convenient to write Eq. (45) as:

$$\sigma_0 \frac{\partial L}{\partial j^\alpha}\frac{\partial j^\alpha}{\partial \bar\phi} \qquad (47)$$

i.e.:

$$\sigma_0 \frac{\partial L}{\partial j^\alpha}\gamma^\alpha\phi \qquad (48)$$

Now the derivative $\frac{\partial L}{\partial j^\alpha}$ for the case of the Lagrangian in Eq. (46) can be found as follows:

$$\frac{\partial L}{\partial j^\alpha} = -k\frac{\partial}{\partial j^\alpha}\left[(j_\beta j^\beta)^{1/2} + j_\beta u^\beta\right]$$

$$= -k\left[\tfrac{1}{2}(j_\lambda j^\lambda)^{-1/2}\left(\frac{\partial j_\beta}{\partial j^\alpha}j^\beta + j_\beta\frac{\partial j^\beta}{\partial j^\alpha}\right) + \frac{\partial j_\beta}{\partial j^\alpha}u^\beta\right]$$

$$= -k\left[\tfrac{1}{2}\frac{1}{\rho_0}(g_{\alpha\beta}j^\beta + j_\beta\delta^\beta_\alpha) + g_{\alpha\beta}u^\beta\right]$$

$$= -k\left(u_\alpha + \frac{j_\alpha}{\rho_0}\right) \qquad (49)$$

Inserting this result back into expression (48) then yields the following for the right hand side of Eq. (43):

$$-\left[\partial_\alpha\frac{\partial}{\partial(\partial_\alpha\bar\phi)} - \frac{\partial}{\partial\bar\phi}\right]\sigma_0 L = -\sigma_0 k\left(u_\alpha + \frac{j_\alpha}{\rho_0}\right)\gamma^\alpha\phi \qquad (50)$$

Finally, combining the results (44) and (50), the overall field equation for the Dirac case is:

$$i\gamma^\alpha \partial_\alpha \phi - m\phi = \sigma_0 k \left( u_\alpha + \frac{j_\alpha}{\rho_0} \right) \gamma^\alpha \phi \tag{51}$$

which completes the derivation of Eq. (8).

**Appendix 2**

The generalised momentum $p^\alpha$ corresponding to the Lagrangian L in Eq. (12) is defined as follows:

$$\begin{aligned}
p^\alpha &= -\frac{\partial L}{\partial u_\alpha} \\
&= k \frac{\partial}{\partial u_\alpha} \left[ \rho_0 (u_\beta u^\beta)^{1/2} + j_\beta u^\beta \right] \\
&= k \left[ \rho_0 \tfrac{1}{2}(u_\lambda u^\lambda)^{-1/2} \left( \frac{\partial u_\beta}{\partial u_\alpha} u^\beta + u_\beta \frac{\partial u^\beta}{\partial u_\alpha} \right) + j_\beta \frac{\partial u^\beta}{\partial u_\alpha} \right] \\
&= k \left[ \rho_0 \times \tfrac{1}{2} \times 1 \times (\delta^\alpha_\beta u^\beta + u_\beta g^{\alpha\beta}) + j_\beta g^{\alpha\beta} \right] \\
&= k(\rho_0 u^\alpha + j^\alpha)
\end{aligned} \tag{52}$$

which establishes Eq. (14).

**Appendix 3**

For a Lagrangian density of the form indicated in Eq. (11):

$$\mathcal{L} = \mathcal{L}_{\text{field}} + \sigma_0 L \tag{53}$$

a general formula for the corresponding energy-momentum tensor exists [9]. This tensor can be expressed naturally in the form:

$$T^{\alpha\beta} = T^{\alpha\beta}_{\text{field}} + T^{\alpha\beta}_{\text{particle}} + T^{\alpha\beta}_{\text{interaction}} \tag{54}$$

where the individual terms are defined to be:

$$T^{\alpha\beta}_{\text{field}} = \left[ (\partial^\alpha \phi) \frac{\partial}{\partial(\partial_\beta \phi)} + (\partial^\alpha \phi^*) \frac{\partial}{\partial(\partial_\beta \phi^*)} - g^{\alpha\beta} \right] \mathcal{L}_{\text{field}} \tag{55}$$

$$T^{\alpha\beta}_{\text{particle}} = \sigma_0 p^\alpha u^\beta \tag{56}$$

$$T^{\alpha\beta}_{\text{interaction}} = \sigma_0 \left[ (\partial^\alpha \phi) \frac{\partial}{\partial(\partial_\beta \phi)} + (\partial^\alpha \phi^*) \frac{\partial}{\partial(\partial_\beta \phi^*)} \right] L \tag{57}$$

For the Dirac case this last term reduces to:

$$T^{\alpha\beta}_{interaction} = 0 \qquad (58)$$

because L is not a function of $\partial_\beta \phi$ or $\partial_\beta \bar\phi$. Also, by using the explicit expression for $\mathcal{L}_{field}$ given in Eq. (6), the term $T^{\alpha\beta}_{field}$ here takes the following specific form:

$$T^{\alpha\beta}_{field} = \left[(\partial^\alpha \phi)\frac{\partial}{\partial(\partial_\beta \phi)} + (\partial^\alpha \bar\phi)\frac{\partial}{\partial(\partial_\beta \bar\phi)} - g^{\alpha\beta}\right]\left\{\tfrac{1}{2}\left[i\bar\phi\gamma^\lambda(\partial_\lambda \phi) - i(\partial_\lambda \bar\phi)\gamma^\lambda\phi\right] - m\bar\phi\phi\right\}$$

$$= \tfrac{1}{2}i\bar\phi\gamma^\lambda \delta^\beta_\lambda (\partial^\alpha \phi) - (\partial^\alpha \bar\phi)\tfrac{1}{2}i\delta^\beta_\lambda \gamma^\lambda \phi - g^{\alpha\beta}\left\{\tfrac{1}{2}\left[\bar\phi(i\gamma^\lambda \partial_\lambda \phi) - (i\partial_\lambda \bar\phi \gamma^\lambda)\phi\right] - m\bar\phi\phi\right\}$$

$$(59)$$

Using Eq. (18) this then becomes:

$$T^{\alpha\beta}_{field} = \tfrac{1}{2}i\left[\bar\phi\gamma^\beta(\partial^\alpha \phi) - (\partial^\alpha \bar\phi)\gamma^\beta \phi\right]$$

$$- g^{\alpha\beta}\left\{\tfrac{1}{2}\bar\phi\left[m\phi - \frac{\sigma_0}{\rho_0}(\partial_\lambda S)\gamma^\lambda \phi\right] - \tfrac{1}{2}\left[-m\bar\phi + \frac{\sigma_0}{\rho_0}(\partial_\lambda S)\bar\phi\gamma^\lambda\right]\phi - m\bar\phi\phi\right\}$$

$$= \tfrac{1}{2}i\left[\bar\phi\gamma^\beta(\partial^\alpha \phi) - (\partial^\alpha \bar\phi)\gamma^\beta \phi\right] + g^{\alpha\beta}\frac{\sigma_0}{\rho_0}(\partial_\lambda S)\bar\phi\gamma^\lambda \phi$$

$$(60)$$

Eqs. (60), (56) and (58) taken together therefore provide the overall energy-momentum tensor corresponding to the Dirac particle/field system considered here.

**Appendix 4:**

The overall 4-divergence of the energy-momentum tensor can be found by considering the field and particle contributions separately. The contribution of the field term is:

$$\partial_\beta T^{\alpha\beta}_{field} = \partial_\beta \left\{\tfrac{1}{2}i\left[\bar\phi\gamma^\beta(\partial^\alpha \phi) - (\partial^\alpha \bar\phi)\gamma^\beta \phi\right] + g^{\alpha\beta}\frac{\sigma_0}{\rho_0}(\partial_\lambda S)\bar\phi\gamma^\lambda \phi\right\}$$

$$= \tfrac{1}{2}(i\partial_\beta \bar\phi\gamma^\beta)(\partial^\alpha \phi) + \tfrac{1}{2}\bar\phi\partial^\alpha(i\gamma^\beta \partial_\beta \phi) - \tfrac{1}{2}\partial^\alpha\left[(i\partial_\beta \bar\phi\gamma^\beta)\right]\phi - \tfrac{1}{2}(\partial^\alpha \bar\phi)(i\gamma^\beta \partial_\beta \phi)$$

$$+ \partial^\alpha\left[\frac{\sigma_0}{\rho_0}(\partial_\lambda S)\bar\phi\gamma^\lambda \phi\right]$$

$$(61)$$

Using Eq. (18), this can be written as:

$$\partial_\beta T^{\alpha\beta}_{\text{field}} = \tfrac{1}{2}\left[-m\overline{\phi} + \frac{\sigma_0}{\rho_0}(\partial_\lambda S)\overline{\phi}\gamma^\lambda\right](\partial^\alpha\phi) + \tfrac{1}{2}\overline{\phi}\partial^\alpha\left[m\phi - \frac{\sigma_0}{\rho_0}(\partial_\lambda S)\gamma^\lambda\phi\right]$$
$$-\tfrac{1}{2}\partial^\alpha\left[-m\overline{\phi} + \frac{\sigma_0}{\rho_0}(\partial_\lambda S)\overline{\phi}\gamma^\lambda\right]\phi - \tfrac{1}{2}(\partial^\alpha\overline{\phi})\left[m\phi - \frac{\sigma_0}{\rho_0}(\partial_\lambda S)\gamma^\lambda\phi\right] + \partial^\alpha\left[\frac{\sigma_0}{\rho_0}(\partial_\lambda S)\overline{\phi}\gamma^\lambda\phi\right]$$
(62)

which cancels to:

$$\partial_\beta T^{\alpha\beta}_{\text{field}} = \tfrac{1}{2}\frac{\sigma_0}{\rho_0}(\partial_\lambda S)\overline{\phi}\gamma^\lambda(\partial^\alpha\phi) - \tfrac{1}{2}\overline{\phi}\partial^\alpha\left[-\frac{\sigma_0}{\rho_0}(\partial_\lambda S)\gamma^\lambda\phi\right]$$
$$-\tfrac{1}{2}\partial^\alpha\left[\frac{\sigma_0}{\rho_0}(\partial_\lambda S)\overline{\phi}\gamma^\lambda\right]\phi + \tfrac{1}{2}(\partial^\alpha\overline{\phi})\frac{\sigma_0}{\rho_0}(\partial_\lambda S)\gamma^\lambda\phi + \partial^\alpha\left[\frac{\sigma_0}{\rho_0}(\partial_\lambda S)\overline{\phi}\gamma^\lambda\phi\right]$$
(63)

Expanding the 2nd, 3rd and 5th terms here then gives:

$$\partial_\beta T^{\alpha\beta}_{\text{field}} = \tfrac{1}{2}\frac{\sigma_0}{\rho_0}(\partial_\lambda S)\overline{\phi}\gamma^\lambda(\partial^\alpha\phi) - \tfrac{1}{2}\overline{\phi}\partial^\alpha\left[\frac{\sigma_0}{\rho_0}(\partial_\lambda S)\right]\gamma^\lambda\phi - \tfrac{1}{2}\overline{\phi}\frac{\sigma_0}{\rho_0}(\partial_\lambda S)\gamma^\lambda(\partial^\alpha\phi) - \tfrac{1}{2}\partial^\alpha\left[\frac{\sigma_0}{\rho_0}(\partial_\lambda S)\right]\overline{\phi}\gamma^\lambda\phi$$
$$-\tfrac{1}{2}\frac{\sigma_0}{\rho_0}(\partial_\lambda S)(\partial^\alpha\overline{\phi})\gamma^\lambda\phi + \tfrac{1}{2}(\partial^\alpha\overline{\phi})\frac{\sigma_0}{\rho_0}(\partial_\lambda S)\gamma^\lambda\phi + \partial^\alpha\left[\frac{\sigma_0}{\rho_0}(\partial_\lambda S)\right]\overline{\phi}\gamma^\lambda\phi + \frac{\sigma_0}{\rho_0}(\partial_\lambda S)\partial^\alpha(\overline{\phi}\gamma^\lambda\phi)$$

with most of these terms now cancelling to yield the result:

$$\partial_\beta T^{\alpha\beta}_{\text{field}} = \frac{\sigma_0}{\rho_0}(\partial_\lambda S)\partial^\alpha j^\lambda \qquad (64)$$

Turning to the particle term, its contribution is:

$$\partial_\beta T^{\alpha\beta}_{\text{particle}} = \partial_\beta(\sigma_0 p^\alpha u^\beta)$$
$$= \sigma_0 u^\beta \partial_\beta p^\alpha + p^\alpha \partial_\beta(\sigma_0 u^\beta) \qquad (65)$$

Assuming the particle's "matter density" is conserved and hence that the continuity equation $\partial_\beta(\sigma_0 u^\beta) = 0$ is satisfied, the 2nd term on the right here will be zero and so Eq. (65) can be expressed in the form:

$$\partial_\beta T^{\alpha\beta}_{\text{particle}} = \sigma_0 \frac{dx^\beta}{d\tau}\frac{\partial p^\alpha}{\partial x^\beta}$$
$$= \sigma_0 \frac{dp^\alpha}{d\tau} \qquad (66)$$

Lagrange's equation of motion for a particle then allows this to be written as:

$$\partial_\beta T^{\alpha\beta}_{particle} = -\sigma_0 \frac{\partial L}{\partial x_\alpha}$$

$$= -\sigma_0 \frac{\partial L}{\partial j^\lambda} \frac{\partial j^\lambda}{\partial x_\alpha} \tag{67}$$

Finally, applying Eqs. (49) and (17), this expression becomes:

$$\partial_\beta T^{\alpha\beta}_{particle} = \sigma_0 k \left( u_\lambda + \frac{j_\lambda}{\rho_0} \right) \partial^\alpha j^\lambda$$

$$= -\frac{\sigma_0}{\rho_0} (\partial_\lambda S) \partial^\alpha j^\lambda \tag{68}$$

It can be seen here that the separate 4-divergences $\partial_\beta T^{\alpha\beta}_{field}$ and $\partial_\beta T^{\alpha\beta}_{particle}$, given in Eqs. (64) and (68), respectively, are not zero, thereby confirming that energy and momentum are being exchanged between the particle and field. Nevertheless the total 4-divergence is zero:

$$\partial_\beta T^{\alpha\beta} = \partial_\beta T^{\alpha\beta}_{field} + \partial_\beta T^{\alpha\beta}_{particle} + \partial_\beta T^{\alpha\beta}_{interaction}$$

$$= \sigma_0 (\partial_\lambda S) \partial^\alpha j^\lambda - \sigma_0 (\partial_\lambda S) \partial^\alpha j^\lambda + 0$$

$$= 0 \tag{69}$$

as required for overall conservation.